# O(p log d) Subgraph Isomorphism using Stigmergic Swarming Agents

H. Van Dyke Parunak[0000-0002-3434-5088]

Parallax Advanced Research, Beavercreek, OH 45431
van.parunak@parallaxresearch.org

**Abstract.** Subgraph isomorphism compares two graphs (sets of nodes joined by edges) to determine whether they contain a common subgraph. Many applications require identifying the subgraph, not just deciding its existence. A particularly common use case, using graphs with labeled nodes, seeks to find instances of a smaller pattern graph with $p$ nodes in the larger data graph with $d$ nodes. The problem is NP-complete, so that naïve solutions are exponential in $p + d$. A wide range of heuristics have been proposed, with the best complexity $O(p^2 d^2)$. This paper outlines ASSIST (Approximate Swarming Subgraph Isomorphism through Stigmergy), inspired by the ant colony optimization approach to the traveling salesperson problem. ASSIST is linearithmic ($O(p \log d)$), and also supports matching problems (such as temporally ordered edges, inexact matches, and missing nodes or edges in the data graph) that frustrate other heuristics.

## 1 Introduction

The maximum partial subgraph isomorphism problem is: given two graphs G1, G2, find the largest subgraph of G1 that is isomorphic to a subgraph of G2. The problem is NP-complete. For example, chemical research frequently needs to find shared structures in large organic molecules, and chemists are among the leaders in research in subgraph isomorphism algorithms [24]. Yet finding a common structure of six atoms between two organic molecules of 50 atoms each requires over $10^{17}$ comparisons.

In addition to molecular design, many other important applications would benefit from the ability to compare even larger graphs. For example:
1. There is increasing use of NOSQL databases such as RDF triple stores, which maintain a network of relationships among millions of data items. Analysts regularly express hypotheses as graphical structures, such as social networks, task networks, and process maps [14], with hundreds of nodes. It would be natural to use the analytical structure directly as a query to see which relations are attested in the data [23].
2. Financial transaction data is a primary data source for detecting financial crime such as money laundering, if it can be searched efficiently for patterns detailing

    known transactional behaviors. But the underlying data is petabytes in volume, with $10^9$ or more nodes.
3. Behavior in computer networks can be characterized by connection graphs derived from IP packet data. A connection is a five-tuple `<source-IP, source-port, dest-IP, dest-port, protocol>`, and an edge joins one connection $C_1$ to another $C_2$ just in case the start time of $C_1$ is earlier than that of $C_2$ and at least one IP address is repeated between them. Such graphs will have millions of connections for reasonable periods of time. Common subgraphs between graphs representing different networks, or the same network at different times, highlight shared behaviors that might indicate common actors.
4. One particular analytic tool is the narrative space [22,26], which fuses many possible causal trajectories. Such graphs, which can contain hundreds of nodes, enable analysts to visualize and interact in both forensic and forecasting problems, but authoring them is time consuming. However, many domains maintain collections of narratives about specific past events [7,13,16,31,32]. Fusing these narratives at their common subgraphs would yield a narrative space for the domain to which they belong, greatly accelerating the analytic process.
5. Fusing patient records in health care can generate a causal model of diagnoses, treatments, and outcomes for resource forecasting and fraud detection.
6. Image recognition makes use of feature graphs that capture adjacency information among different features. More efficient subgraph isomorphism algorithms would allow powerful new image search capabilities.

In these cases, and others that could be added, the graphs in question range from $10^3$ to $10^6$ nodes or even more. With even the 50-node graphs of organic chemistry posing computational limits, a breakthrough is clearly needed to address these problems.

This paper describes ASSIST (Approximate Swarming Subgraph Isomorphism through Stigmergy), a novel heuristic inspired by ant colony optimization (ACO) [8]. This class of algorithms is inspired by social insects such as ants and termites, who coordinate their work by depositing chemicals, called "pheromones," in the environment and making their decisions based on the current pheromone strengths in their vicinity. These chemicals evaporate over time, and those that are reinforced by many ants converge to the selected solution. In nature, these mechanisms enable termites to construct mounds with separate floors and rooms, and ventilation systems to exhaust waste gasses. They enable ants to construct minimal spanning trees joining their nests to food sources. ACO algorithms replace the ants with simple software agents, the pheromones with increments that the agents make to variables on the different locations that they visit, and evaporation with a periodic attenuation of the accumulated pheromone by a specified percentage. The results have proven very successful in other highly complex problems such as the traveling salesperson. ASSIST applies these techniques to subgraph isomorphism.

Section 2 surveys related work. Section 3 describes the swarming approach to the traveling salesperson problem (TSP) that inspires ASSIST. Section 4 describes the basic ASSIST algorithm, and Section 5 shows how it supports extensions of the problem that frustrate other heuristics. Section 6 concludes.

## 2  Related Work

We compare ASSIST first with other graph matching algorithms, then with other stigmergic systems.

Graph matching is an active research area. Convenient surveys include [4,5,10,24]. We situate ASSIST in this context along five dimensions, highlighting how its characteristics address the four key challenges of Accuracy, Scalability, Robustness, and Accessibility (understandable by non-technical users). Table 1 summarizes the impact of these dimensions on the challenges.

*Table 1: Characteristics of ASSIST compared with other methods for subgraph isomorphism*

| Dimension (ASSIST in **Bold Italics**) | Benefits of ASSIST |
|---|---|
| Exact vs. ***Approximate*** | Scalability, Robustness |
| Transformed vs. ***Direct*** | Robustness, Accessibility |
| Deterministic vs. ***Stochastic*** | Accuracy |
| Single vs. ***Population*** | Robustness, Accessibility |
| Entire vs. ***Incremental*** | Scalability |

**Exact** methods (such as Ullman's pioneering algorithm [30]) are guaranteed to find matching subgraphs, but are combinatorially prohibitive (because the problem is NP complete), so they do not *scale* to large problems, and they are not *robust* to noise in the pattern or data. Successive filtering methods [19] apply a sequence of exact methods to the data, hoping to reduce its size before attempting to identify subgraphs, but are stymied by incompletely labeled data that makes early filters ineffective. ASSIST belongs to the family of **approximate** methods, as do the studies cited in the rest of this section.

Spectral methods [17,34], Estimation of Distribution Algorithms (EDAs [2]), and Optimal Transport approaches [35] (with complexity $O(p^2d^2)$) **transform** the problem (into matrix parameters, probabilistic graphical models, or probability distributions, respectively). Matrix methods are not *robust* to poorly conditioned graphs, while transformations in general are less *accessible* to analysts. ASSIST (and many other methods) manipulate the pattern and data graphs **directly**.

Most methods are **deterministic**, giving the same answer each time they are run, but such approaches are vulnerable to falling into local optima, compromising *accuracy*. ASSIST (like EDAs [2] and GAs [6,33]) is **stochastic**, allowing it to escape from local optima, and also permitting a dynamic trade-off between probability of detection and false $p_d$ and FAR.

Most methods produce a **single matching**, which is best by some criterion internal to the algorithm, but may not be *robust* to corrupted patterns or data. EDAs [2] and GAs [6,33] consider a **population** of individual competing solutions, allowing alternative matches to emerge, but typically focus down to a single solution, so that analysts never see alternatives. ASSIST also considers multiple solutions, but ranks them probabilistically (by the strength of the pheromone field on each one), enhancing *accessibility*.

Most methods reason about the **entire** pattern as a whole, but some, like ASSIST, are **incremental** [12,25,28,29], starting with local matches within each of the two graphs and expanding the match. Incremental match construction enhances *scalability* by limiting the effort spent on candidate matches that end up failing.

Stigmergic swarming, the heart of ASSIST, has been applied successfully in many areas, including the traveling salesperson problem [27], telecommunications routing [15], and our use of it to outperform game-theoretic, Bayesian, and human forecasters in geospatial reasoning [21]. ASSIST is an aggressive extension of these techniques. Unlike geospatial reasoning (but like routing problems) it is not limited to the regular structure of a lattice, but handles arbitrary graphs. Unlike previous graph applications, it has separate pheromone families for nodes and edges and we will extend it to deal with time-sequenced graphs to handle time compactly. In addition, its use of a third pheromone family to implement quorum sensing mechanism is unparalleled in previous applications of stigmergy.

## 3  Approach: Ant Colony Optimization

One of the most successful applications of ACO is to the traveling salesperson problem: given a set of points in two dimensions and a road network among them, find the shortest Hamiltonian circuit (visiting every node exactly once and returning to the start). This problem is of great importance in problems such as logistics [11] (optimizing fuel usage for truck fleets), telecommunications [3], and wiring plans for printed circuit boards [1], among others, and is successful enough to find widespread commercial use (for example, www.antoptima.com).

In these applications, successive swarms of software agents (digital ants) explore alternative tours in parallel, choosing randomly at each step among the unvisited vertices, but weighting its choice to favor edges with the strongest pheromones left by previous waves, and then returning home when all vertices have been visited. After completing a circuit, each ant deposits pheromone on the edges it has traversed, with strength inversely proportional to the length of the overall path. After each wave of agents explores the graph, all pheromone strengths evaporate at rate $\rho$. Edges that are part of shorter circuits accumulate more pheromone, and attract more agents, while edges that are not visited evaporate, and over time a highly competitive path emerges (Figure 1). Depending on details of the application, the time complexity is $O((n \log n)/\rho)$ [20].

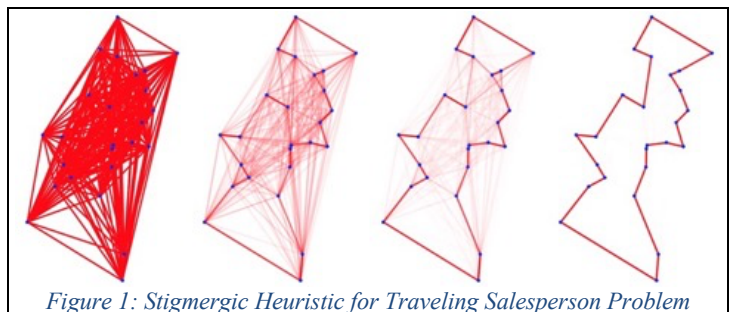

*Figure 1: Stigmergic Heuristic for Traveling Salesperson Problem*

ASSIST is inspired by this algorithm. Instead of Hamiltonian circuits, we ask our ants to look for paths between the pattern and data graphs that traverse matching fragments of both graphs (typically, a single edge). As the pheromones from many ants stabilize, a high quality matching emerges.

## 4   The ASSIST Heuristic

ASSIST uses stigmergic software agents that travel over and between the graphs being compared. "Stigmergic" means that the agents interact with one another, not by direct messages, but by leaving markers inspired by insect pheromones (in this case, scalar fields) in their environment (the graphs they are exploring). The environment manipulates fields by aggregating deposits of the same type by different agents, evaporating them over time, and propagating them to neighboring locations. Each agent increments one or more of the fields at its current location, senses the fields in its immediate neighborhood, and then moves based on a weighted combination of the field strengths at different accessible locations.

We begin by identifying *peers*, that is, nodes in one graph with the same labels or other characteristics as nodes in the other. These are matching subgraphs of size 1. Figure 2 illustrates the peers in the data of node A in the pattern. If there were another node A in the pattern, it would also be peered with the two nodes labeled A in the data.

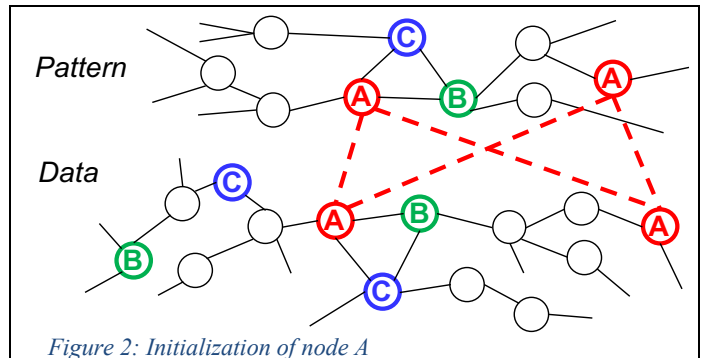

*Figure 2: Initialization of node A*

We reasonably assume that the database with the data has a balanced index, so the time requirement for peering is $O(\log d)$, where $d$ is the number of nodes in the data. Since we match each of the $p$ nodes in the pattern, the overall complexity is $O(p \log d)$, or linearithmic. Other steps in the algorithm scale linearly in the number of nodes, but need to be iterated to convergence, so the overall complexity is $O(p \log d + Cp)$, with a larger contribution from the linear component ($C > 1$).

In addition to peering, initializing involves having each node and edge propagate its label (or other characteristic critical for matching) to its neighborhood. Each node and edge maintains a vector of how strong each label is in its vicinity, allowing an agent on one node to estimate a gradient for a given label type from that node to one of its neighbors, and thus search for a path to the desired label or characteristic.

Our fundamental insight is that matching subgraphs are made up of matching edges, which can be discovered locally by swarming agents that seek for neighbors of already-matched nodes and reinforce the pheromone levels on those nodes and the edges that join them. Nodes that participate in more than one shared edge are further reinforced. Multiple edges in a shared subgraph reinforce each other, and the larger the subgraph, the more pheromone its nodes and edges accumulate. Pheromone on nodes in either graph that do not have shared edges eventually evaporates to 0.

Each swarming agent starts on a peered node in the pattern graph and seeks a path corresponding to the dashed loop in Figure 3. It can move *between* the two graphs to peered nodes (black arrows 1, 3), and (guided by propagation of digital pheromone) *within* a graph to adjacent nodes (purple arrows 2, 4). As it seeks such a cycle, the agent maintains a history of the nodes and edges it has visited, with their labels and other properties, and can draw on this information as well as the pheromone fields it encounters in making movement decisions. This stateful reasoning will be the key to respecting temporal order in transactions. Exploring such a path is a constant-time operation, and can be pursued by many agents in parallel. Propagation within each graph communicates to each node the labels and relative weights of its neighbors, while deposits between graphs allow shared edges to reinforce one another's strength. Propagation can extend to higher-order neighbors as well, allowing the agents to detect approximate matches where a node in one graph is missing or mislabeled in the other.

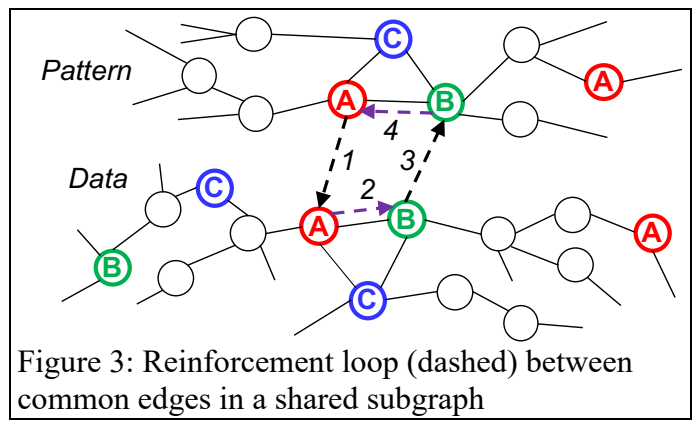

Figure 3: Reinforcement loop (dashed) between common edges in a shared subgraph

This core approach regularly finds the kernel, but does not reliably find the *maximal* common subgraph). The problem is that the detection loop finds only a single link, and reinforces its endpoints equally whether they are part of a small subgraph or a large one.

Our approach to this problem is inspired by quorum sensing in multicell aggregates such as slime molds and biofilms [9,18]. In the biological domain, each cell emits a chemical, senses the aggregate level of the chemical, and switches from individual to group-mode behavior when the aggregate exceeds a threshold. We use a similar mechanism to tell nodes in a subgraph the size of the subgraph of which they are a part. We add a node-based quorum pheromone field augmented by all nodes linked by edges with non-zero edge pheromone. When an ant completes a cycle, it updates node and edge pheromone inversely proportional to the length of the cycle it has found. Then each node propagates quorum pheromone along all edges with non-zero edge pheromone, so nodes in larger subgraphs accrue more quorum pheromone. This information can modulate the operation of the system in various ways.

We detect termination in this system when the sum of the pheromones in each graph differs from one step to the next by less than a specified amount (currently $10^{-6}$). The resulting pheromone field gives a probabilistic measure of the similarity of subgraphs in the two graphs. Because it is probabilistic, and not all or nothing, it can handle approximate matchings and missing or extraneous data. In addition, it lends itself to a visual user interface, highlighting relevant regions of the world view for the analyst.

Figure 4 shows the evolution of the pheromone field on the nodes in two graphs in the current prototype. Initially, we place pheromone on all peered nodes (matching subgraphs of size 1), As the swarm discovers matching edges and reinforces pheromone on their endpoints, and as other pheromone evaporates, only the matching subgraphs remain. In this example, the graphs were seeded with the triangular subgraph ABC, but the algorithm discovered that in both graphs randomly grown from this seed, node S was adjacent to A, and included it in the result. (The edge from A to S in the lower graph is present, but obscured by C.)

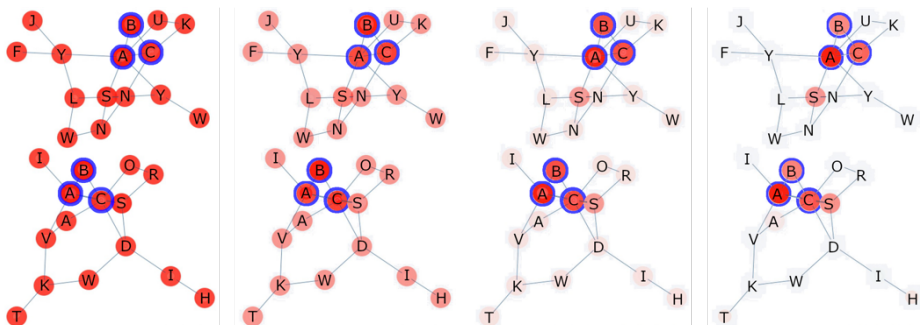

*Figure 4: Behavior of ASSIST Prototype*

## 5   Extensions

Figure 3 shows the basic matching mechanism. But some problems have additional complexity. For example:

a.  Node labels might not match exactly. A data graph might have a personal name, but a pattern might have simply the designation "person." Maintaining an ontology to allow such abstractions is not difficult, but the simple matching mechanism described above would miss the match.
b.  The graph might be directed. For example, in a graph of financial transactions, the edges indicate transactions, and the movement of money from A to B is not the same as movement from B to A. This directedness imposes a time ordering on the edges, which may be recorded either as clock time or as a partial order over the edges.
c.  A node or edge might be missing entirely in either the pattern or the data.

These complications frustrate many other subgraph algorithms, but ASSIST can accommodate them.

The pattern in Figure 3, repeated in Figure 5 (Imprecise), handles case a (here, matching 'B' with 'b') by allowing the ants to consult an ontology in case of mismatch to see if one of the nodes subsumes the other. The total pheromone deposited in the case of an imprecise match will be less than that deposited for an exact match.

Figure 5 (Temporal) handles temporal matches by asking the agent two crosses two edges in the data graph before returning to the pattern, remembering the sequence of these edges, and then seeking a sequence of edges in the pattern with the same order to return home.

Figure 5 (Missing) handles missing data using pheromone propagation across multiple edges. An agent can then sense the presence of an otherwise desirable node not immediately adjacent to its current node and move to it. In this case, as in the case of imprecise data, the pheromone deposited at the end of the circuit will be less than in the case of a perfect match.

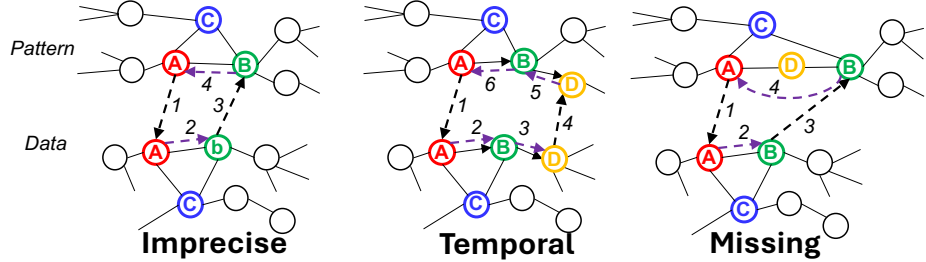

*Figure 5: ASSIST can handle more complex matches*

## 6    Conclusion

ASSIST is a revolutionary approach to subgraph isomorphism inspired by the computational mechanisms used by social insects. In addition to being much faster than competing approaches, it is (like the insects who inspire it) robust and adaptable, able to accommodate noisy, irregular data.